\documentclass[aps,prb,reprint,floatfix,superscriptaddress,showpacs,letterpaper,longbibliography]{revtex4-2}

\usepackage{graphicx}
\usepackage{bm}
\usepackage{amsmath}
\usepackage{amsfonts}
\usepackage{amssymb}
\usepackage{epstopdf}
\usepackage{xfrac}

\begin{document}

\title{On neutron holography,  neutron interferometry and  neutron orbital angular momentum } 

\author{Wolfgang Treimer}

\author{Frank Haußer}
\affiliation{Department of Mathematics,  Physics and Chemistry Berliner Hochschule f\"{u}r Technik, D-13353 Berlin, Germany}

\author{Ingeborg Beckers}
\affiliation{Department of Mathematics,  Physics and Chemistry Berliner Hochschule f\"{u}r Technik, D-13353 Berlin, Germany}

\author{Martin Suda}
\affiliation{Security and Communication Technologies, Center for Digital Safety and Security, AIT  Austrian Institute of Technology GmbH, A-1210 Vienna, Austria}

\author{Terrence Jach}
\affiliation{Material Measurement Laboratory, National Institute of Standards and Technology, 100 Bureau Drive, Gaithersburg, Maryland 20899, USA}


\begin{abstract}
 
A neutron Laue crystal interferometer has been reported by Saranac $ \it{et~al.}$ to demonstrate neutron holography of a spiral phase plate. Using its two coherent beams as the object and reference beams, the resulting interference pattern was interpreted as a hologram.  This interference pattern was then reported to reconstruct neutron beams with various intrinsic orbital angular momenta. There are serious doubts about the method for generating neutron orbital angular momentum with a crystal interferometer.  Due to the extremely different lateral coherence lengths in the interferometer, one should expect the pattern described as a hologram to be asymmetric.  In addition, a neutron crystal  interferometer always produces a Moiré image  where the beams are combined in the final crystal, that appears as a one-dimensional enlargement of the interference pattern. We present computer simulations showing that the images presented as holograms  can be computed as conventional interference patterns assuming the phase shifts of ordinary neutrons passing through objects  located in one or the other beam path of the interferometer. Additionally, we discuss the complications of using a crystal interferometer for holography, while raising the question of whether the intrinsic orbital angular momentum states of neutrons are necessary or sufficient to explain the recorded images.

\end{abstract}
\maketitle

\vspace{2pc}

\setlength\parindent{0pt}

\section{Introduction}
Experiments with photons often serve as a predecessor for analogous experiments and findings in neutron optics. 
Photons, characterized by spatially varying amplitude and phase distributions, can possess intrinsic orbital angular momentum (OAM) in units of $\hbar$, which was shown theoretically in 1992 \cite{Allen1992,Barnett1994}, and demonstrated experimentally in various ways (see e.g. \cite{Rubinsztein2017}) .
The successful demonstration of quantized OAM for photons has opened up avenues for applications in quantum entanglement, quantum information science, and imaging \cite{Mair2001}. 
For example, the orbital angular momentum of light beams was used early on to rotate microscopic particles trapped in optical tweezers \cite{He1995}.  Recent advancements have enabled the generation of X-rays with orbital angular momentum using a free-electron laser oscillator \cite{Huang2021}.  A comprehensive study of optical vortices, which includes OAM manipulations from topological charge to multiple singularities, can be found e.g. in \cite{Yao2011,Shen2019}. The creation of OAM states with photons was soon extended to electrons \cite{Bliokh}. The field of electron orbital angular momentum presents promising opportunities \cite{McMorran2011}.\\

Neutrons described as a wave packet with a similar spatially varying amplitude and phase distribution should also possess an orbital angular momentum (n-OAM) in addition to their spin.  By analogy with photons and electrons, this should occur by passing a coherent neutron wave through a spiral-shaped phase object. The interaction should produce a helical wavefront, resulting in intrinsic n-OAM. 
However, the very description of a helical phase front implies a coherence of phase across the phase object \cite{Allen2011,Yao2011}.  Realizing this and conducting further investigations of neutron angular momentum and its corresponding applications are therefore of great interest.  \\

The first attempt to demonstrate a twisted phase front was performed in 2015 within a neutron Laue crystal interferometer \cite{Clark}. In this experiment, macroscopic spiral phase plates of varying thicknesses were inserted in one path of the  interferometer, resulting in interferograms that were attributed to n-OAM. However, objections have been raised that the lateral coherence lengths of the neutron wave used were far too small to generate n-OAM in the manner described \cite{Cappelletti2018, Jach2022}. The observation is more correctly explained as ordinary neutrons with a small transverse coherence length that experience a simple phase shift in traversing a given path length of matter. When interfering with their alternate path in the interferometer, each neutron traces out a small element of the phase object as a whole. When coarsely imaged, the interference pattern is indistinguishable from what might be expected from a macroscopically broad n-OAM state.  Furthermore, it has been argued that the use of the crystal interferometer and the corresponding Moiré imaging led to a misinterpretation of these interferograms  \cite{TreimerarXiv2023, Treimer2024}.  \\

OAM states of photons and electrons have also been produced in diffraction by grating patterns containing a dislocation \cite {Yao2011,Bliokh}.   Gratings of this type have been produced holographically by the interference of a vortex beam of photons (an OAM state) with a reference beam in an optical interferometer. It is therefore tempting to posit that the use of a phase plate of the type described in \cite{Clark}, placed in one leg of a Laue crystal interferometer, combined with a deflecting wedge in the other arm of the interferometer, produces a "neutron hologram" \cite{Sarenac2016}, analogous to what might be produced with light in a Mach-Zender (MZ) interferometer \cite{Leith1963}. The purpose of this paper is to analyze the manner in which these two situations differ to the extent that they are not analogous, and the results cannot be used to justify the creation of an n-OAM state.  \\

\section{General Considerations}
We begin by stating some objections to a series of assumptions in Ref.  \onlinecite{Sarenac2016} that are problematic regarding holography in a neutron interferometer: 
\begin{itemize}

\item The neutron interferometer based on Laue diffraction does not process neutrons in the same way as photons in an optical MZ interferometer.

\item The Fresnel diffraction of an object that gets recorded in a hologram is not satisfied by the conditions of the Laue interferometer.

\item Because of the 2-d nature of the Laue interferometer, the essentially 3-d nature of a helical neutron wave function cannot be captured - only the conventional scalar phase of an ordinary neutron is captured. If one captured the 3-d nature of the helical neutron, we believe one would observe the effects of scattering outside of the 2-d scattering plane, which is not the case. 

\item The addition of a prism or wedge in the reference leg of the neutron crystal interferometer does not act to produce a reference beam as it does in optical holography. The function of the prism to bend the beam of the optical holography demonstration and the function of the prism to add simple scalar phase to the neutron in one arm of the interferometer have no similarity of function. 

\item The interference that is occurring between the paths of the neutrons in the two legs of the neutron interferometer, unlike that of an optical holography experiment, is occurring due to the much smaller wavelength of the neutron, typically on a sub-nanometer scale that cannot be recorded. The "interference pattern" observed in the neutron interferometer is really the overlap of two nanometer scale interference patterns by two branches (in each leg) of the dispersion surface (see Section III), which differ only slightly in wavevector. As a result, it is really a Moiré pattern caused by the overlap that is recorded on a physically observable scale.
\end{itemize}

\begin{figure}[h]
\centering
\includegraphics[width=6cm]{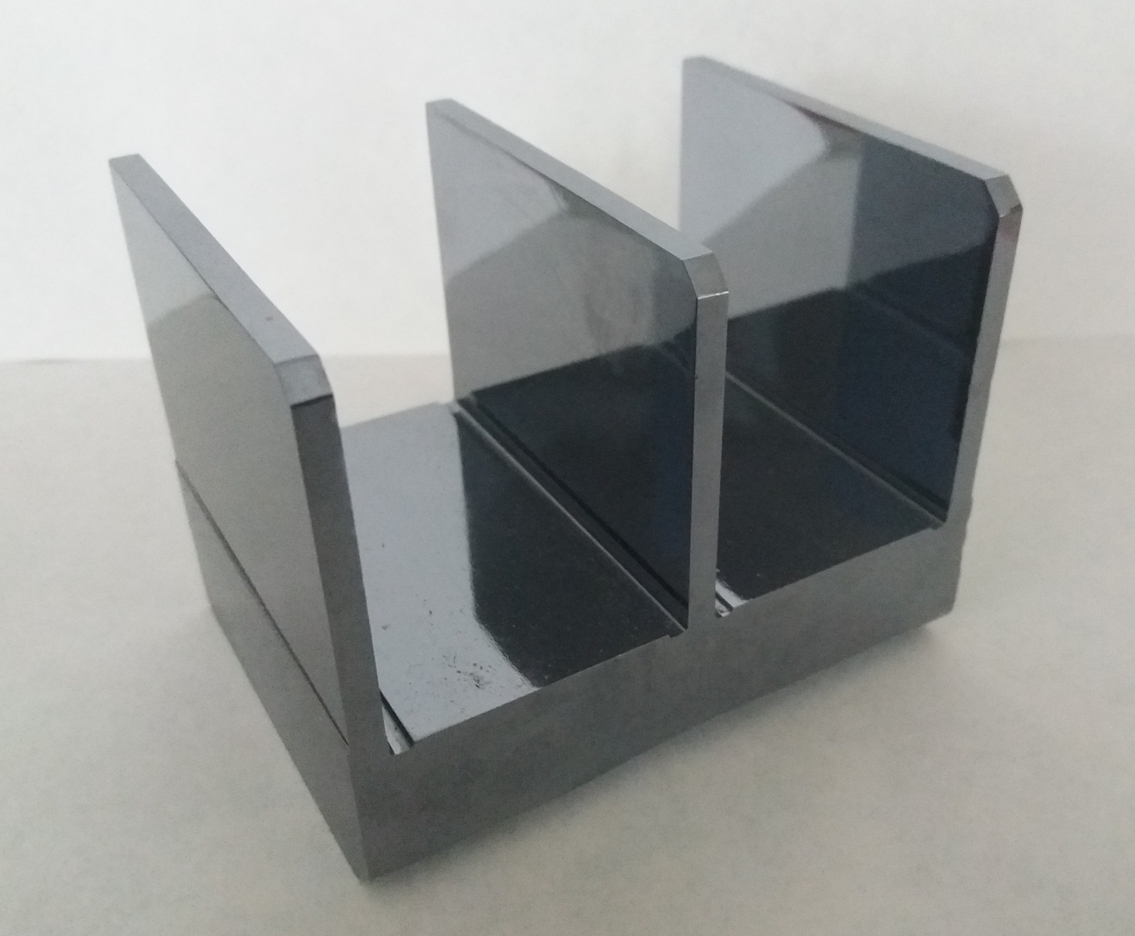} 
\caption{A Si single crystal neutron interferometer as was used in  \cite{RauchTreimerBonse1974, TreimerThesis}, length, height and width were (67.8, 60.0, 60.0)[mm], each plate thickness = 4.46mm, plate  distances  = 27.23mm, (220) reflection lattice planes.  }
\label{fig:Si-Interferometer}
\end{figure}

The paper is organized as follows: In Section III the dynamical theory of the Laue neutron interferometer is described. It briefly summarizes the interaction of thermal neutrons with a perfect crystal and within a perfect crystal  neutron interferometer 
(Fig.\ref{fig:Si-Interferometer}). For all mathematical details, see  \cite{RauchSuda1974,RauchPetrascheck1978,RauchPetrascheck1984, Petrascheck1988, RauchWerner2000, Suda2006, SudaSkript, LemmelSoftware2022}. In section IV neutron OAM and holography are discussed. In Section V the calculation of interferograms and computer simulations are presented and in Section VI conclusions are drawn.

\section{Neutron interferometry}

For the interaction of a neutron wave with a perfect crystal lattice, described by the stationary Schr\"{o}dinger Equation, the wave function $\psi(\vec r)$ is assumed to have a periodic interaction with $\delta$-function potentials $V(\vec r)$ [22]:

\begin{equation}
\begin{aligned}
({\nabla ^2} + k^2)\psi (\vec r) &= \frac{{2m}}{{{\hbar ^2}}}V(\vec r)\psi (\vec r),\\
V(\vec r)  &= \frac{{2\pi {\hbar ^2}}}{m}{b_c}\sum\limits_{i,j} {\delta (\vec r - {{\vec r}_{i,j}}) }\\
\end{aligned}
\label{Eq.:Schrödinger}
\end{equation} 

Here, $k=\frac{2\pi}{\lambda}$ is the vacuum wave number, $\lambda=\frac{h}{mv}$ is the wave length, $m$ is the neutron mass, $v$ is the neutron velocity, $h=2\pi\hbar$ is Planck's constant, and $b_{c}$ is the coherent scattering length. $\vec r_{i,j}=(\vec r_{j}+\vec r_{i})$ is the vector of lattice position to elementary cell position $\vec r_{j}$ and to atom position $\vec r_{i}$ in the cell under consideration. 

\begin{figure}[hbtp]
		\centering
		\includegraphics[width=7cm]{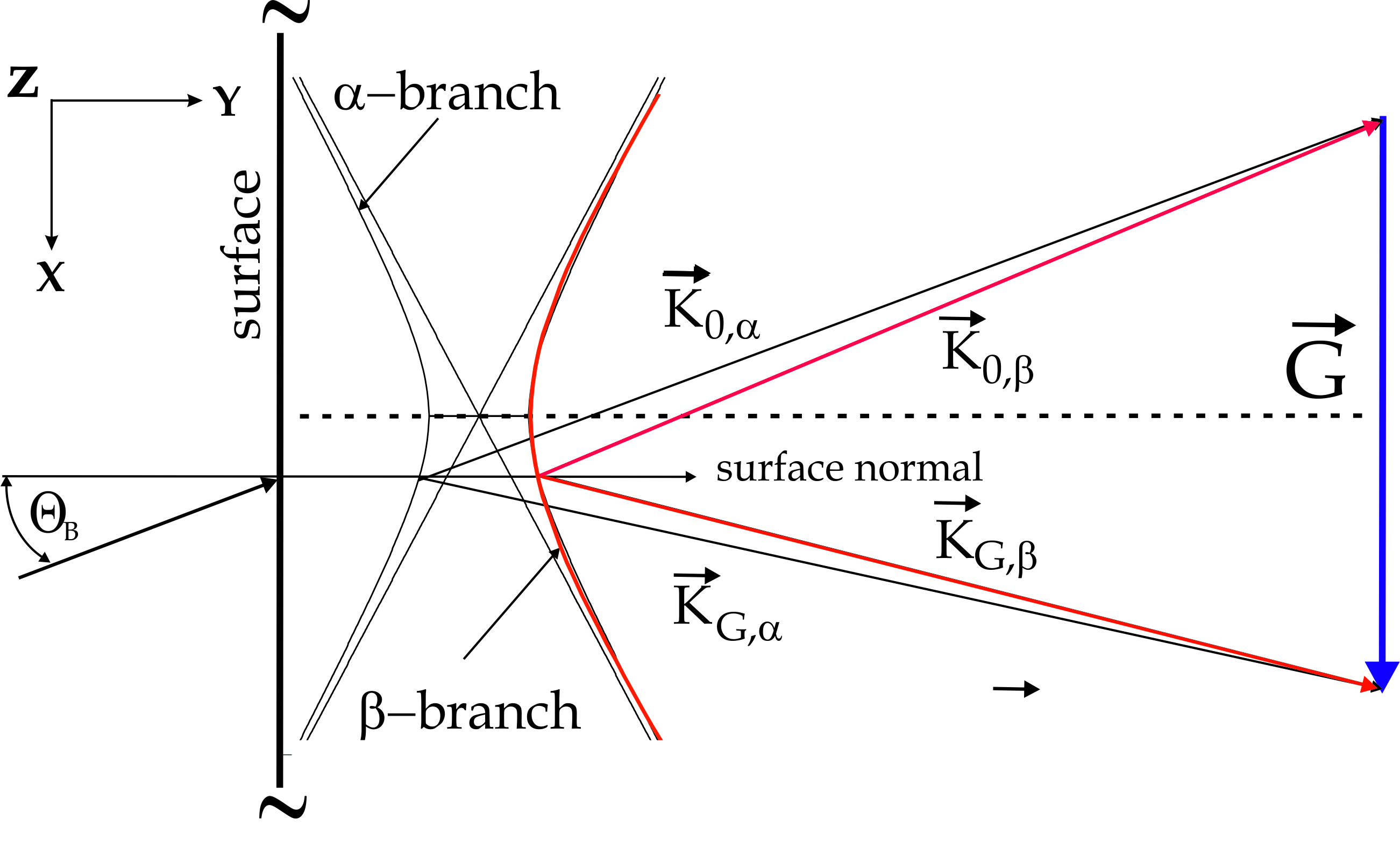}
		\caption{Dispersion surface for  the Laue case diffraction, rays emerging from the $ \alpha $ branch or $ \beta $  branch into O and G directions are independent solutions of the  Schrödinger Equation, i.e.  ${\psi _{1,2}} = \psi _O^{1,2} + \psi _G^{1,2} $,  {$\vec K$ are the wavevectors in the crystal.} }
		\label{fig:dispersion surface}
\end{figure}

The Equation above yields, in the case of dynamical Laue-diffraction, two independent solutions, $\psi_{1,2}=\psi_{O}^{1,2}+\psi_{G}^{1,2}$, {where} $"1"$ stands for the $\alpha$ branch and $"2"$ for the $\beta$ branch, respectively (Fig.~\ref{fig:dispersion surface}).  $\vec G$ is the reciprocal lattice vector. The rays for Laue diffraction in a perfect lattice are shown in Fig.~\ref{fig:dispersion surface}. That means, that waves stemming from the $\alpha$ branch traveling in the $O$ $and$ $G $ directions are $one$ solution of Eq.(\ref{Eq.:Schrödinger}) and waves stemming from the $\beta$ branch traveling in the $O$ $and$ $G $ directions are the $other$ independent solution of Eq.(\ref{Eq.:Schrödinger}).
The waves from different dispersion branches $\alpha$ and $\beta$ (Fig.~\ref{fig:dispersion surface}) can be collected into a forward beam and a diffracted beam. Behind the crystal the forward beam becomes 
\begin{equation}
\begin{aligned}
\frac{\psi_0(D)}{\psi_e} &= \frac{X_2 e^{i k \epsilon_1 D / \cos(\gamma)} - X_1 e^{i k \epsilon_2 D / \cos(\gamma)}}{X_2 - X_1}\\
 &= v_0,   \quad \text{with }  \cos (\gamma ) = \frac{{{k_ \bot }}}{k} 
\end{aligned}
\end{equation}
The diffracted beam  behind the crystal is  given by $ \psi_{G}(D) $:
\begin{equation}
\begin{aligned}
 \frac{\psi_G(D)}{\psi_e} &= \frac{X_1 X_2 \left[ e^{i k \epsilon_1 D / \cos(\gamma)} - e^{i k \epsilon_2 D / \cos(\gamma)} \right]}{X_2 - X_1} e^{i \vec{G} \cdot \vec{r}}\\
 & = v_G \cdot e^{i \vec{G} \cdot \vec{r}} 
\end{aligned}
\end{equation}
The following quantities are defined here: {$X_{1,2}=u_{1,2}(\vec G)/u_{1,2}(0)$, where $u_{1,2}(0)$ are the amplitudes of the two wavefields in forward direction and $u_{1,2}(\vec G)$ are the amplitudes of the two wavefields in the diffracted direction,} $D$ is the crystal thickness, and  $\psi_{e}$ defines the incoming plane wave, of wavevector $k$ whose component normal to the surface is $k_\bot$. $\epsilon_{1,2}$ are the two excitation errors of the two wave fields.
That means that $(k\epsilon_1)$ and $(k\epsilon_2)$ are the differences between internal wavevectors of the dispersion surfaces ($\alpha$ and $\beta$) and the crossing points of $\vec{k}_0$ and $\vec{k}_G$, which are the vacuum wave vectors of the forward and the diffracted direction.
$v_{0}$ and $v_{G}$ are called crystal functions.
{The waves in both the O and G directions are coherent sums of $\alpha$ and $\beta$ branch contributions (see Fig.~2).}
The proof of these coherent properties was given by C.G. Shull in 1968, in which he demonstrated so-called neutron pendulum solutions using Laue reflections for different wave lengths which can be explained only with the coherence of waves from the $\alpha$ and $\beta$ branch [27]. 
\\\\
Therefore, one can write $\psi_{1,2}=\psi_{O}^{1,2}+\psi_{G}^{1,2}$, where $1$ and $2$ denote the branches $1=\alpha$ and $2=\beta$ of the dispersion surface. Thus the corresponding intensities behind the crystal become
\begin{subequations}
\label{fields} 
\begin{eqnarray}
\left|\frac{\psi_{G}(D)}{\psi_{e}}\right|^{2} &= &\frac{\sin^{2}(A_{1}\sqrt{1+y^{2}})}{1+y^{2}},\label{equationa}\\
\left|\frac{\psi_{0}(D)}{\psi_{e}}\right|^{2} &=& 1 -
\frac{\sin^{2}(A_{1}\sqrt{1+y^{2}})}{1+y^{2}},\label{equationb}
\end{eqnarray}
\end{subequations}

where
\begin {equation}
 {A_1}=\frac{k}{{2 \cos (\gamma )}}\frac{{V(\vec G)}}{E} D, 
\end {equation}\\
 is proportional to the crystal thickness $D$, $V(\vec G)$ is the Fourier transform of $V(\vec r)$, $E$ is the energy of the neutron) and $y\approx -2\sin(\Theta_{B})\delta \Theta_{B}$ denotes the deviation from the Bragg angle $\Theta_{B}$.  Each dispersion branch $\alpha$ and $\beta$ generates waves in both the $O$  and the $G$ directions inside the crystal as shown in Fig.~\ref{fig:dispersion surface}\\

\begin{figure}[h]
		\centering
\includegraphics[width = 7cm]{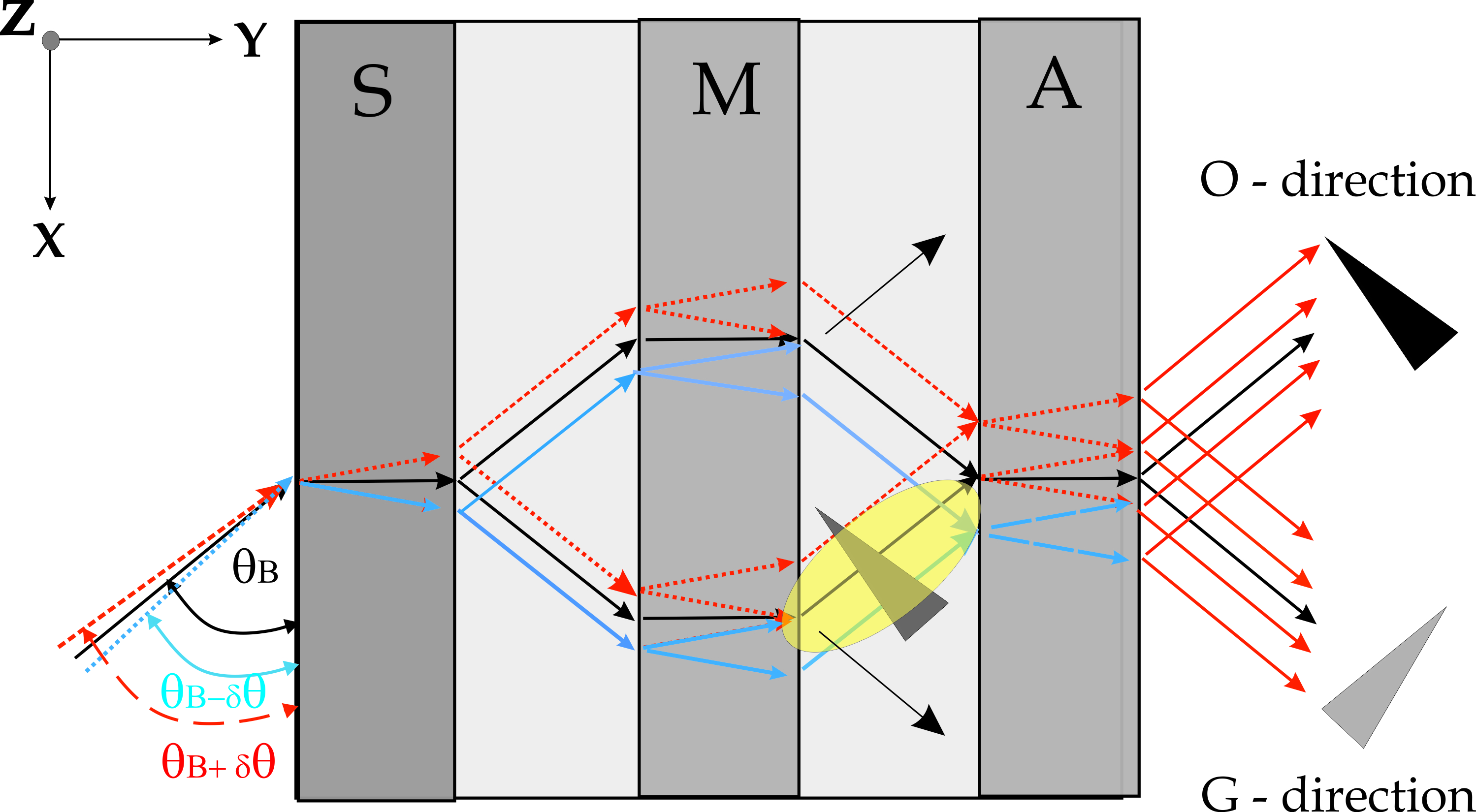}
		\caption{Rays in interferometer, S, M and A are the splitter plate, mirror plate and analyser plate. $\psi_e$ is the incident neutron wave. Angular deviations from $ \theta_{B} $ cause wave fields belonging to $ \alpha  $ and $ \beta $ branch of the dispersion surface (Fig.~\ref{fig:dispersion surface}) and result in a spreading of neutron flux known as the Borrmann-Delta (Fig.~\ref{fig:Borrmann-Delta}).  (The thickness of the crystal plates is increased so that the rays in the plates become more visible). Upper path is denoted by $(I)$ and lower path is denoted by $(II)$.}
		\label{fig:Rays in interferometer}
		\end{figure}

For neutron interferometry one has to determine wave functions for three successive Laue diffractions as shown in Fig.~\ref{fig:Rays in interferometer}. 
{Behind the mirror crystal (M), the two rays converge and interfere either before or within the third crystal plate (A).} We can denote the upper path by $(I)$ and the lower one by $(II)$. The coherent waves arising from Laue diffraction in all three crystals via path I and path II belong to a scattering plane defined by the incident wave vector of the neutron wave $ \vec{k} $ and the reciprocal lattice vector $ \vec{G} $ of the first crystal plate, here in the \{x,y\} plane (Fig.~\ref{fig:Rays in interferometer}). \\

Following [22], behind the interferometer and depending from a phase shift $\chi$ the intensity in the $O$ direction becomes ($\psi_{0}^{I}=\psi_{0}^{II}$)
\begin{eqnarray}
I_{0}(\chi)=|\,\psi_{0}^{I}\,+\psi_{0}^{II}\,e^{-i\chi\,}|^{2}=\frac{1}{2}|\psi_{0}|^{2}\,[\,1+\cos(\chi)\,]
\end{eqnarray}
realizing that 
\begin{eqnarray}
I_{0}(\chi)+I_{G}(\chi)=|\psi_{0}|^{2}+|\psi_{G}|^{2}=const.
\end{eqnarray}
Thus $I_{G}(\chi)$ is oscillating in opposite phase to $I_{0}(\chi)$.\\

For this approach we reduced the calculations to an (infinite thin) $(x,y)$- scattering plane where coherent contributions of off-scattering planes are omitted. This is justified by an incident Gaussian wave packet $\Gamma(k-k_{0})$: 
\begin{equation}
\begin{array}{l}
\Gamma (k - {k_0}) = \frac{1}{{{{[2\pi {{(\delta k)}^2}]}^{1/4}}}} \cdot {\mkern 1mu} \exp \left[ { - \frac{{{{(k/{k_0} - 1)}^2}}}{{4{{(\delta k/{k_0})}^2}}}]} \right],\\
\psi (x,t) = \frac{1}{{\sqrt {2\pi } }}\int\limits_{ - \infty }^\infty  {\Gamma (k - {k_0}) \cdot {\mkern 1mu} \exp [{\mkern 1mu} i{\mkern 1mu} (kx - \Omega t){\mkern 1mu} ]{\mkern 1mu} dk} 
\end{array}
\label{Eq.: wavepacket}
\end{equation}
%
where $\Omega=\hbar{}k^2/(2m)$, $(\delta{}k/k_{0})=\sigma_{x}$ and $k_{0}$ is the mean wave number. The mean square deviation $\sigma_{x} $  ($\sim \mu $m) is about two orders of magnitude greater than $\sigma_{z}$ ($ \sim $ nm). Therefore a one-dimensional wave packet is used. 
Another argument for a reduction of a three-dimensional to a one-dimensional wave packet here is that any contribution due to out-of-plane scattering would be registered in the same detector pixel (100$\mu$m x 100$\mu$m  in \cite{Sarenac2016}) as the contribution originating from the scattering plane. \\

Thus, any interference pattern can only arise from the one-to-one interference between the wave from path I and the wave from path II (see Fig.~\ref{fig:Rays in interferometer}), and it occurs exclusively within their scattering plane, forming a 'pair of coherent beams' \cite{BonseHart1965}. Note, that there is only one neutron in the interferometer at a time and only self-interference occurs. \\

Two other inherent effects of dynamic diffraction are characteristic of Laue crystal interferometry. The first one is the angular amplification of the neutron flux in the Borrmann Delta (Fig.~\ref{fig:Borrmann-Delta}), which arises from small deviations in the incident wave from the exact Bragg angle or from slight variations in the incident wavelength.  {For neutrons with a  wave length $ \lambda$  and  $ \Delta\lambda/\lambda\ll$10$^{-4}$ the whole Borrmann Delta is always excited. If any other  neutron is incident at  exactly the same direction but at a different wave length, the energy flow is split within the Borrmann Delta and  exits  at "a" ($ O^{*} < a < A$)  and "b", ($ O^{*} < b < B $)   as seen in Fig.~\ref{fig:Borrmann-Delta} \cite{Petrascheck1988,Shull1968} . \\
\begin{figure}[hbtp]
		\centering
		\includegraphics[height=5cm]{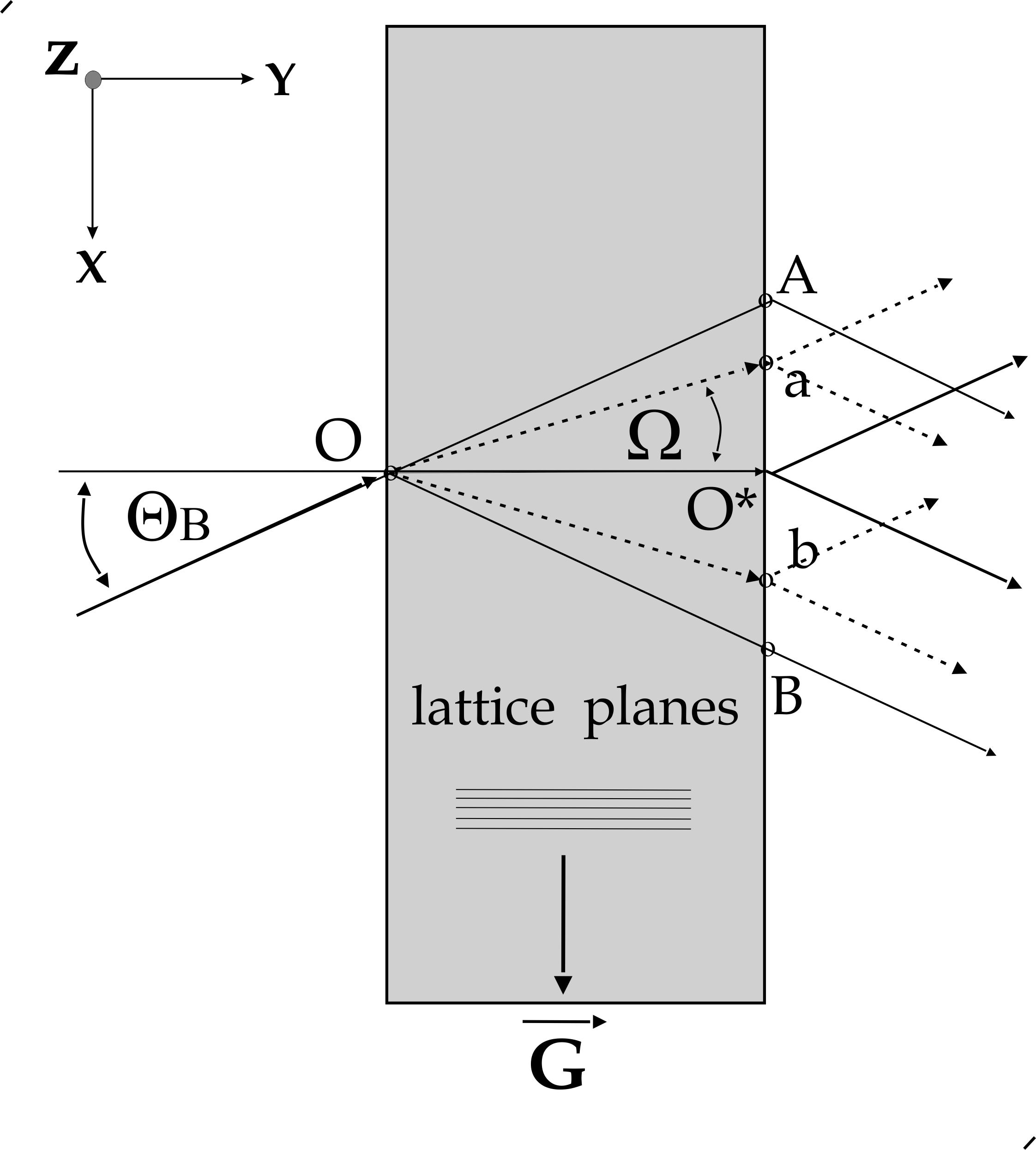}
		\caption{A neutron beam incident at the Bragg angle $\Theta_B$ is coherently split into the forward direction $\widehat{OA}$ (O-beam) and the diffraction direction $\widehat{OB}$ (G-beam), just as waves from "a" and  "b".   $\widehat{OAB}$ is called the Borrmann Delta. }
		\label{fig:Borrmann-Delta}
\end{figure}

The second  inherent effect in neutron interferometry  is  the magnification of any interference pattern  due to Moiré imaging \cite{Bonse1969}.
The Laue solution for a plane crystal plate results in a standing wave pattern behind the first crystal plate (i.e., the interferometer beam splitter S) creating two coherent waves each, both in direction O and  in G. (Fig.~\ref{fig:Rays in interferometer}, Fig.~\ref{fig:Borrmann-Delta}).
From the middle crystal plate (M, mirror) two waves converge  again due to Laue-diffraction and overlap in front (or in) the third crystal plate (A, analyser) and form an interference pattern whose shape depends on the optical path difference between path I and path II. 
The resulting fringe pattern still has the same fringe spacing as the Bragg planes on a sub nanometer scale \cite{Bonse1969}. This fringe pattern excites wave fields in the analyser crystal A and, through the superposition of the Bragg planes of the analyzer crystal A, it produces a   pattern with infinitely magnified fringe spacing when the fringe spacing equals the lattice spacing,  in reality typically by a factor of  10$ ^{7} $ - 10$ ^{8} $  compared to the original interference pattern's fringe spacing. \\

Thus the detected intensity pattern is not periodic in the actual wavelength of the neutrons, which would not, in fact, be observable. The magnification occurs  in the scattering plane only,  since the standing interference pattern behind each crystal plate is always parallel to the reflecting lattice \{x,y\} planes. By shifting the third crystal plate (A) parallel to  ($ \vec G $ ) by half a crystal lattice spacing relative to the interference pattern, the Moiré  pattern changes,  from maximum to minimum. The magnification, in addition to proving the correctness of the dynamical theory for the entire interferometer, enables measurements of displacements on an{ \AA}ngstrom scale \cite{Bonse1969} which are visible on a length scale that is vastly expanded.\\

At this point, it is useful to compare a Mach-Zender light interferometer  with a crystal interferometer of the MZ type, which appears to share some similarities in geometry. In  both systems, the phase of one or both paths can be manipulated relative to the other. This manipulation is achieved by a phase-shifting object, which is characterized by a so-called $ \lambda $-thickness ${D_\lambda } $.  ${D_\lambda } $ is the thickness of a material that shifts the phase of  the wave  by 2$ \pi$. For neutrons  ${D_\lambda } = 2\pi /( N {b_c} \lambda ) $, ($ b_{c} $ is the neutron coherent scattering length, N is the number of atoms/unit volume, $\lambda$  is the neutron wave length). \\  

An MZ light interferometer accepts radiation from nearly all angles of incidence, allowing these to contribute to the interference. As a result, interference can always be observed, provided that the optical path lengths in arms I and II coincide within the longitudinal coherence of the source. The final interference pattern can be observed directly.   The Laue diffraction version of this interferometer only accepts and transmits neutrons in an extremely narrow angular range about the Bragg angle. Each neutron produces  \textit{two}  solutions in \textit{each} leg of the interferometer, resulting in a complicated wavefield at the output.  Furthermore, the interference pattern observed is not, in fact, periodic in the wavelength of the neutrons but is a Moiré pattern highly magnified in one dimension along the scattering plane. These differences between an MZ  light interferometer and a crystal interferometer will indicate impediments to true neutron holography.}

\section{coherence, neutron OAM, and holography}

Perfectly helical wave fronts imply complete radial spatial coherence across the beam cross-section, a condition for orbital angular momentum \cite{Allen2011,Yao2011,Huang2021,McMorran2011}. The coherence of the incident light beam in an optical MZ interferometer is determined by the source (e.g., a laser) before entering the interferometer. Because of the bosonic character of laser photons, the entire beam can be considered to have transverse coherence, so the presence of a phase device (i.e. a spiral phase plate, SPP) can create an orbital angular momentum state that is the diameter of the entire beam.  However, it has been demonstrated that as the transverse coherence of the laser beam is reduced to a diameter much smaller than the dimensions of the spiral phase plate, the transmitted photons are no longer in OAM states but trace out small individual areas of the phase plate so that the integrated image appears unchanged\cite{Cappelletti2020}. \\

Because of the fermionic character of the neutrons, the transverse coherence is limited to what is achievable with a single wave packet and does not extend beyond that. In this sense, the azimuthal phase is imposed only on those neutrons (wave packets) that lie within their transverse coherence from the spiral axis of the phase plate. \cite{Bliokh}\\  


In contrast to the coherence  properties of laser beams used in optical OAM experiments, neutron beams have small and sometimes different  coherence lengths in the two transverse directions, typically of the order of nanometers to micrometers, as well as low intensity. Detailed experiments concerning  effective transverse coherence of neutrons  can be found e.g. in \cite{RauchWerner2000,Majkrzak2014}. In small-angle neutron scattering (SANS), axis-symmetric beam collimation ensures that the lateral coherence lengths are equal,  so that homogeneous illumination of particles from nanometer to as large as micrometer dimensions is guaranteed. Recently, the first experiments involving small-angle neutron scattering and neutron orbital angular momentum states have been reported  \cite{Sarenac2024}.\\

However, it is different in perfect crystal diffraction  because only one lateral coherence length comes into play,  which lies in the scattering plane, i.e.  parallel to the $x$-axis and  parallel to the crystal surface  (Fig~\ref{fig:Borrmann-Delta}). This was demonstrated, for instance, by slit diffraction \cite{Shull1969} and phase grating experiments performed  with a high-angle resolution double crystal diffractometer where a lateral coherence length $ \sigma_{x}$  of up to 100 µm could be achieved \cite{Treimer2006, Ebrahimi2010,Treimer2011}. Measurable, coherent contributions from scattered waves \emph{outside} the scattering plane were not detected. This is obvious, because the lateral coherence lengths $ \sigma_{l} = \sigma_{x,z} $ differ considerably,  i.e.  $ \sigma_{z} \ll  \sigma_{x} $ \cite{RauchWerner2000}.\\
 
 In the Laue crystal interferometer, the transverse coherence properties are determined  by both the preparation of the incident neutron beam and the Laue reflections occurring in the three crystals.  Only a region near the axis of a phase plate between approximately 60 nm and a few micrometers, corresponding to radially different lateral coherence lengths in $x$ and $y$ of the neutron wave packet, as created in the crystal interferometer, could be effective. \\

Therefore, the assumption for the transverse wave function  $ \psi_{t} $, 
\begin{equation}
\psi_t = (2\pi\sigma^2)^{-1/2} e^{-\frac{(x-x_0)^2 + (y-y_0)^2}{4\sigma^2} }
\label{Eq.:Pis_t}
\end{equation}
 in \cite{Sarenac2016},  (this paper $ z \equiv y $) where the coherence lengths, $ \sigma_x = \sigma_y \equiv \sigma $  were set equal,  cannot be correctly assumed.  The coherence lengths in the crystal neutron interferometer that was employed have been measured to differ by about two orders of magnitude:  $ \sigma_z  \approx$  80nm $\ll \sigma_x \approx 5\mu m $,  see e.g. \cite{Pushin2008}. \\

\begin{figure}[hbtp]
\centering
\includegraphics[width=7cm]{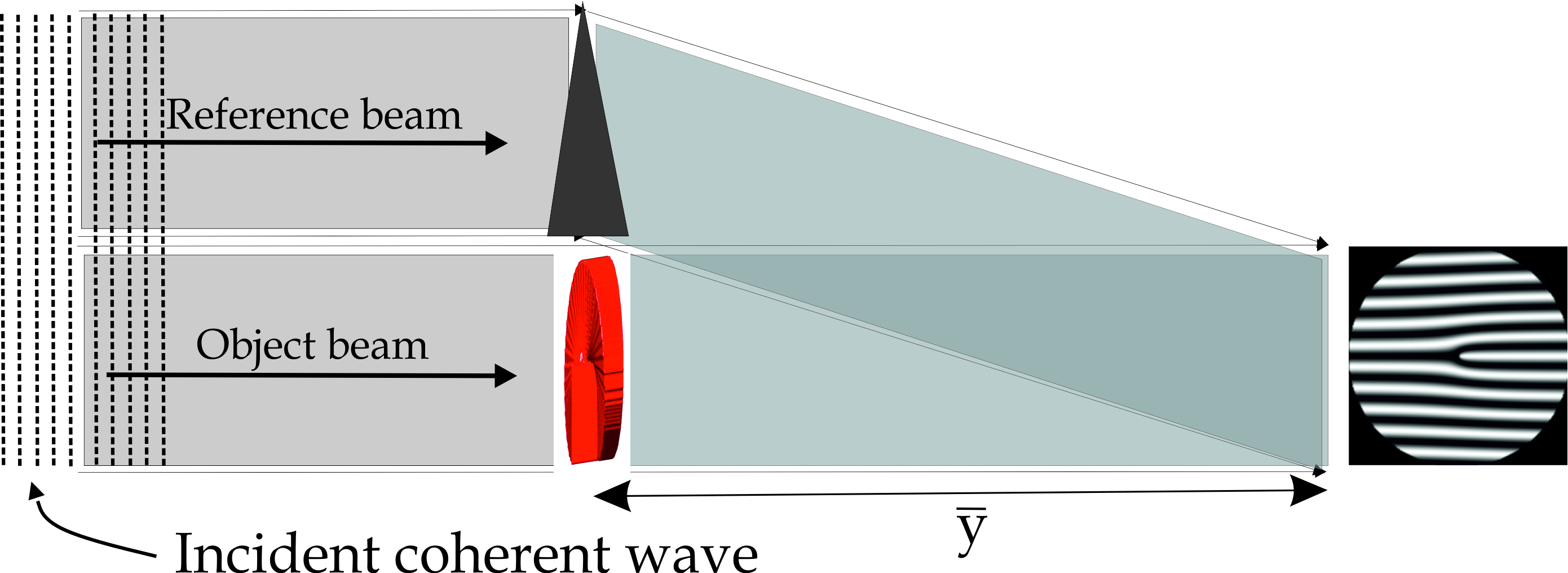}
\caption{The experimental setup by Leith and Upatnieks is similar to that shown in Fig.~1 of \cite{Leith1963}, where a wedge was placed in the reference beam and a transparent image in the object beam. In \cite{Sarenac2016}, a refraction wedge was also used in the reference beam, with an SPP serving as the object in the object beam.
The coherent superposition of the optical wavefronts modulated by the refractive wedge and the SPP yields a fork-like interference pattern.}
\label{fig:LeithUpatniek}
\end{figure}
\begin{figure}[hbtp]
\centering
\includegraphics[width=8cm]{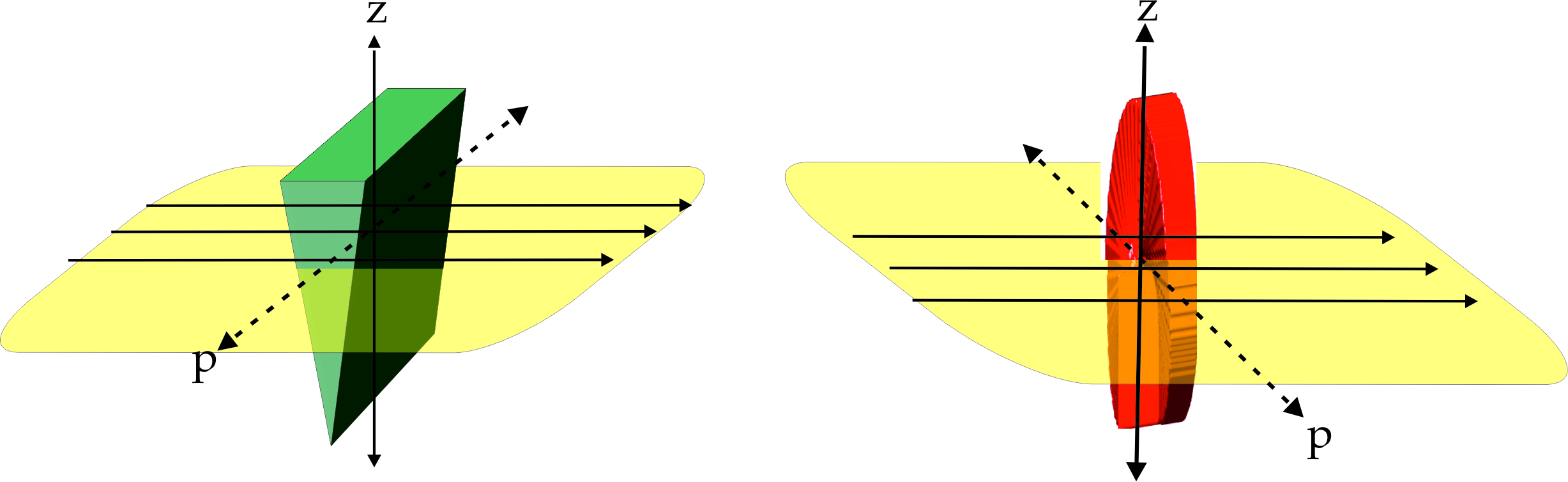}
\caption{The wedge and spiral phase plate (SPP), as used as phase-shifting elements in the neutron interferometer in \cite{Sarenac2016}. }
 \label{Keil und  Spirale Strahlen}
 \end{figure}

A second factor that is necessary for genuine holography, Fresnel diffraction,  is not met in the neutron interferometer, as explained below.
The experiment described in \cite{Sarenac2016} is an adaptation of an  experiment in 1963 by Leith and Upatnieks \cite{{Leith1963}} (see Fig.~\ref{fig:LeithUpatniek}). There,  an incident monochromatic wavefront was split into an object beam and a reference beam using wave front splitting. 
A prism in the reference beam deviates this portion of the beam and superimposes it at a distance $  \overline{y} $  with the object beam, which itself creates a Fresnel diffraction pattern:  "The superposition of the reference and object beams results in a fringe pattern that is overlaid on the Fresnel diffraction pattern of the object  and a photographic plate records the resultant pattern, thereby producing the hologram" \cite{Leith1963}.  The function of the  wedge was  to deviate the beam, but it also  served to imprint location-dependent phase information using the interference pattern with the object beam.
In \cite{Sarenac2016}, it was attempted to realize this experiment (off-axis method of optical holography) with a neutron crystal interferometer, using a wedge and an SPP as object, indicated in Fig.~\ref{Keil und  Spirale Strahlen}.
  
\begin{figure}[hbtp]
\centering
\includegraphics[width=8.5cm]{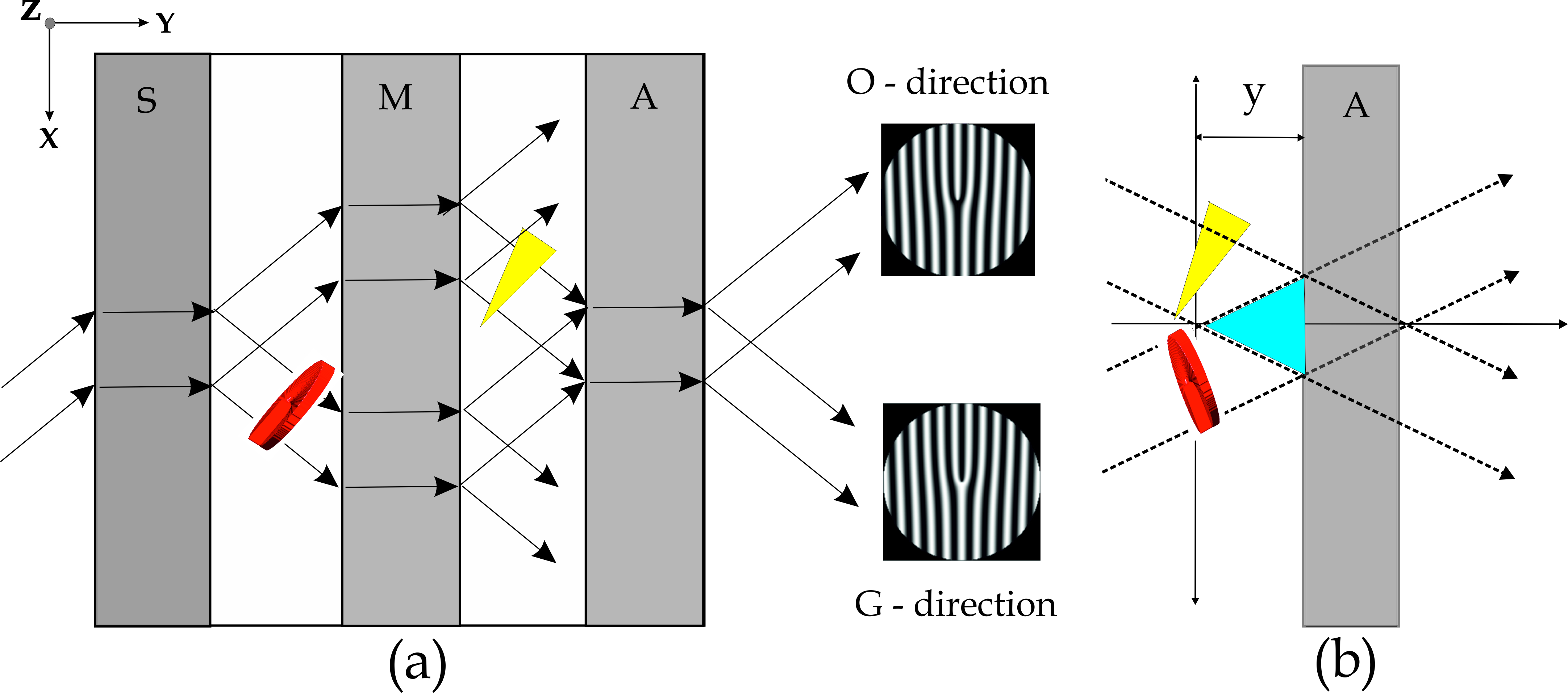} 
\caption{(a) Rays and phase objects in the neutron interferometer as given in  \cite{Sarenac2016}:  The SPP (red object) was placed in the lower beam path and the wedge (yellow object) in the upper one. Their transmitted waves interfere with each other in front of or in the analyzer crystal (A) and form an interference pattern which, when superimposed on the crystal lattice of (A), produces a Moiré pattern. 
The interferograms calculated in \cite{Sarenac2016}  are shown in the O and G directions. Note, the interference patterns were calculated as the SPP and the horizontal wedge are shown in Fig.~1(b) in  \cite{Sarenac2016}, see text. The Fresnel number F for this SPP position is 0.295.  
(b) Beam geometry when a wedge (yellow object) and the SPP (red object)  are  closest to the analyzer crystal (A).  $y$ is now the minimum distance of the phase objects  from (A), resulting in a Fresnel number F = 0.857 $< 1 $, i.e. one has still Fraunhofer diffraction.} 
\label{fig:IFM+wedge+SPP+Interferogram}
\end{figure}

Off-axis optical holography, as used in \cite{Leith1963} (here Fig.~\ref{fig:LeithUpatniek}), splits a coherent beam (via wavefront division) into an object beam, which illuminates the object, and a reference beam.
Using a neutron interferometer, the first crystal plate splits the incoming beam into two coherent partial beams (via amplitude splitting) by Laue diffraction  (see Fig.~\ref{fig:IFM+wedge+SPP+Interferogram}). The SPP was placed in one beam path,  the wedge was  in the other beam path, assumed to serve as the reference beam, similar to the setup in the Leith-Upatnieks experiment   (Fig.~1 in \cite{Leith1963}, here Fig.~\ref{fig:LeithUpatniek}). 
But there, the wedge was  used to deviate the reference beam in order to be superimposed with the object beam  (see Fig.~\ref{fig:LeithUpatniek}) and it  served also to imprint location-dependent phase information on the interference pattern of the object beam as cited above. This requires the superposition of both beams, spatially resolved on the photo-plate (2D detector).

 The Fresnel diffraction by the object modulates the fringe pattern of the wedge. 
Therefore, the  wedge as used in \cite{Sarenac2016}  did not  have the same function as in the Leith-Upatnieks experiment for several reasons:  

First, in the case of a neutron interferometer, the so-called 'reference beam' does not have to be deflected in order to be superimposed with the 'object beam', because the diffraction at the second plate (mirror M) automatically deflects  the beam toward the third crystal plate. 

Second, the wedge used (and calculated) in \cite{Sarenac2016} was positioned as in Fig.~\ref{Keil und Spirale Strahlen}, which deflected the ray upwards (or downwards) by approx. \( 5 \cdot 10^{-7} \)rad, and not as shown in Fig.~1(b) in \cite{Sarenac2016}, i.e. the wavefront was deflected in the wrong direction, and not as the interference patterns show. Consequently, the wavefront deviation in \cite{Sarenac2016}  caused by the wedge had no measurable influence on the observed interference pattern for simple geometrical reasons. With a pixel size of $100\, \mathrm{\mu m \times 100\, \mu m}$, each wavefront deviation hit the same detector pixel. In the neutron interferometer, the wedge merely served as an additional phase shift element.\\

As previously mentioned, in holography, the observed pattern results from the superposition of the Fresnel diffraction from the object and the interference pattern generated by the reference beam. Fresnel diffraction or near field diffraction of an object requires its  monochromatic (spatially) coherent illumination \cite{Leith1964}.
Near-field diffraction can be characterized by F, the so-called  Fresnel number,  and  curved  wave fronts  with $ \theta \approx a/ y $,  where  $ \theta  $  is the angle made by the wave from the aperture  'a' ( i.e. the source of coherent radiation)  and  $ y $  is the  short distance of the object from the screen (or two-dimensional detector).
In order to distinguish Fresnel diffraction from Fraunhofer diffraction, the so-called Fresnel number $ F = \frac{a^2}{\lambda y} $  must satisfy the condition $ F \geq 1 $,  with  $ \lambda  $ being the wavelength of the incident beam.   \\

For the experiment in \cite{Sarenac2016} the Fresnel number F can be estimated as follows. 
Using Fig.~1(b) in  \cite{Sarenac2016} the SPP was placed in the neutron interferometer between crystal plates (S) and (M),  much more than 5cm from crystal plate (A) (see (Fig.~\ref{fig:IFM+wedge+SPP+Interferogram}(a)). 
The crystal plate (A) is the closest entity that interacts with the interference pattern.
Assuming  a  maximum coherent area 'a'   = 2µm,  $ \lambda = 2.71\cdot10^{-10} $ m, a SPP diameter 15 mm and y at least 5 cm apart from the crystal plate (A), one produces  a Fresnel number $F = 0.295  \ll  1$. 
However, in the neutron interferometer used, the minimum possible distance $y$ is limited as \( y_{\text{min}} = 17 \ \text{mm} \)  (see Fig.~\ref{fig:IFM+wedge+SPP+Interferogram}(b)), which  is the minimum distance that the SPP and the wedge can maintain from analyzer plate A without their interference patterns overlapping. This results in a Fresnel number of \( F = 0.857 \), which is less than 1, indicating the regime of Fraunhofer diffraction.

\begin{figure}[hbtp]
\centering
\includegraphics[width=6cm]{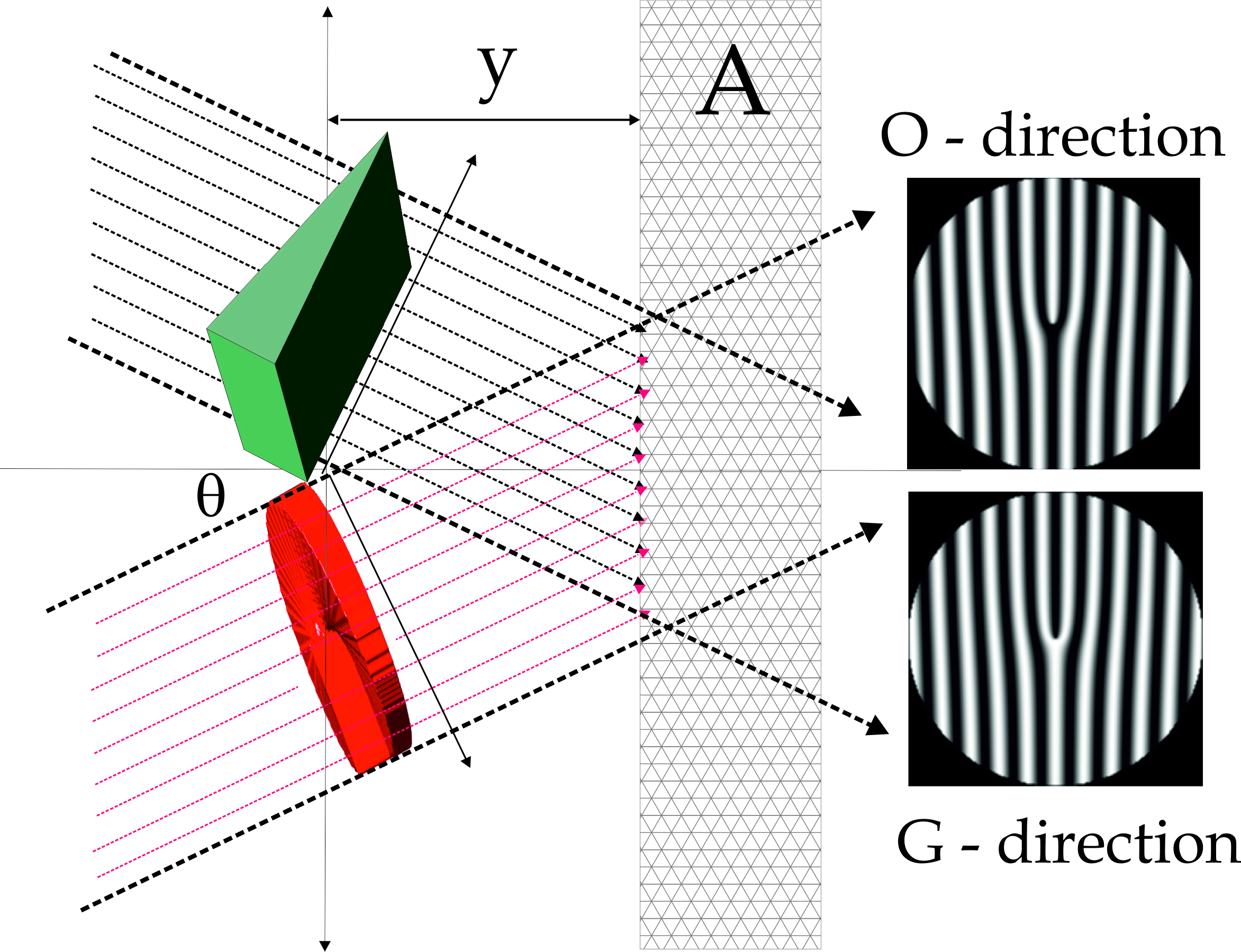}
\caption{Interference pattern with wedge and SPP,  $y$ is  the minimum distance of both phase objects  from the analyzer crystal.  Note the complementary images of O -  and G - beam}
\label{fig:Pattern in O and G-direction detail}
\end{figure}

 {As mentioned above} in the experiments in \cite{Sarenac2016} the wedge was positioned vertically, not as shown in Fig.~1(b) in \cite{Sarenac2016}. The measured and calculated fringe pattern there in Fig.~3(b) in  \cite{Sarenac2016} cannot be caused by a horizontal wedge.  The resulting interference pattern in O - and G - direction would look as shown in Fig.~\ref{fig:IFM+wedge+SPP+Interferogram},  details of the calculated interference pattern are shown in Fig.~\ref{fig:Pattern in O and G-direction detail}.

\section{Calculation of interferograms and computer simulations}

{In a neutron interferometer, the wave packet is highly asymmetrical,} allowing Eq.(\ref{Eq.: wavepacket})  to be used for simulating the interference patterns. Another assumption can be made due to the small value of  $ \sigma_{x} $  (a few $\mu  $m) in comparison to the detector resolution of 100 µm in the x-direction. This discrepancy causes the interference effects to be effectively canceled at each detector pixel.
 The same applies  to  $\sigma_{y} $,  particularly because  $ \sigma_{y} \ll \sigma_{x} $ and $ \sigma_{y} \ll 100\,\mu \mathrm{m}$. This significantly simplifies the overall calculations.
We calculated  the two-dimensional phase front of the SPP which was placed in one (lower) beam path (II)  and  two-dimensional phase front  of the wedge (WE) in the other (upper)  beam path (I)  using the Radon transform RT  \cite{Radon1917}  for the angle of incidence $\vartheta = 90^{o} $, i.e. perpendicular to the SPP and to the wedge (Fig.~\ref{fig:IFM+wedge+SPP+Interferogram}). 

\begin{figure}[h]
\centering
\includegraphics[width=8cm]{{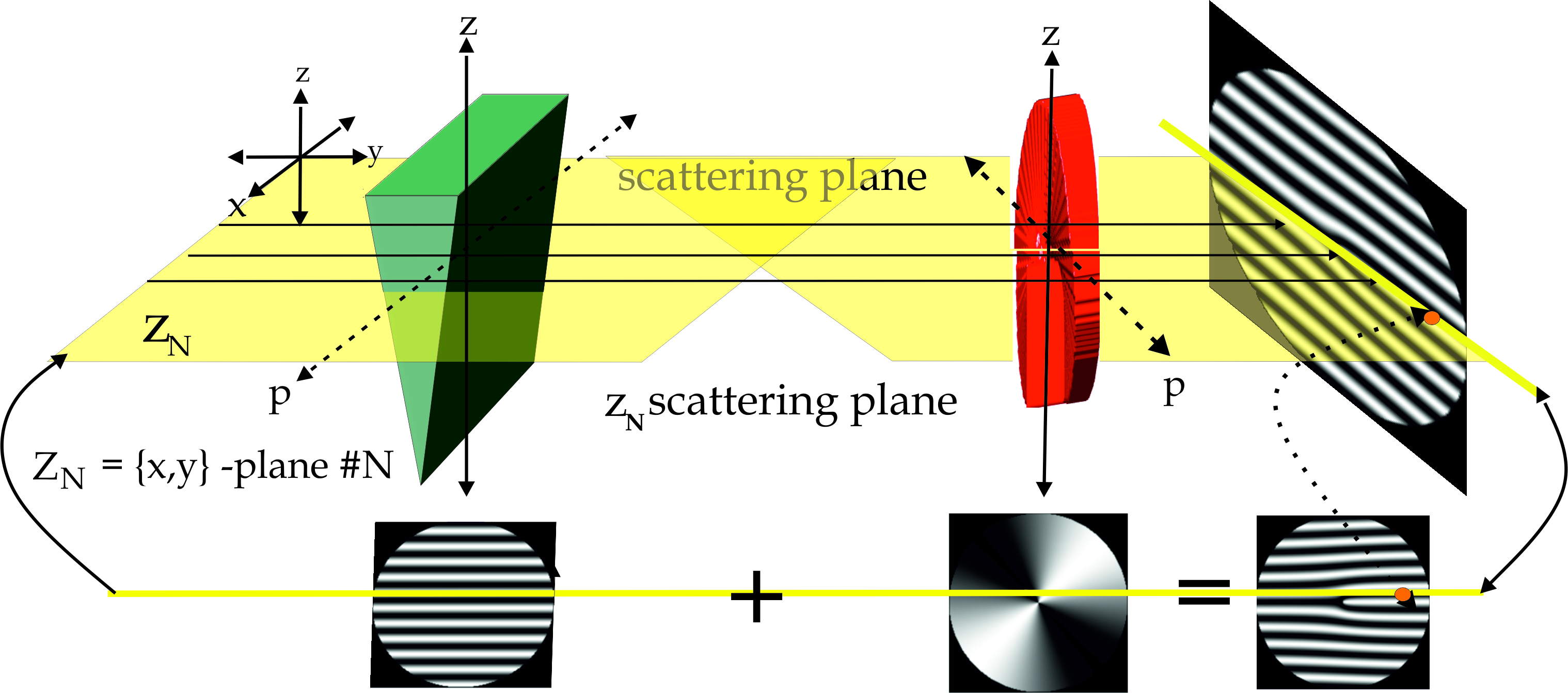} } 
\caption{Each phase object was divided into N  slices parallel to the $ \{x,y\} $ plane, $ z_N $.  The line at the bottom shows the addition of patterns.} 
\label{fig:interfernce pattern as sum}
\end{figure}

Each phase object  (see Fig.~\ref{Keil und  Spirale Strahlen})  was divided into N slices in $z$-direction, i.e. the  $RT\{SPP\} $ and  $RT\{WE\} $ which  provides the spatially dependent transmission lengths of each partial beam in the $y$-direction of all  $(x,y)$-planes (Fig.~\ref{fig:interfernce pattern as sum}).

 For each slice WE$ _{z} $ and SPP$_z $, the corresponding Radon Tranform  was calculated as 
\begin{subequations}
\label{RT} 
\begin{eqnarray}
{\mathrm{RT}}\{ \mathrm{SPP}_z \} (p,\vartheta)  &= \int\limits_{ - \infty }^\infty  {\int\limits_{ - \infty }^\infty  {\mathrm{SPP}_z(x,y) \cdot \hat \delta  \,dx\,dy} } \\
{\mathrm{RT}}\{ \mathrm{WE}_{z} \} (p,\vartheta)  &= \int\limits_{ - \infty }^\infty  {\int\limits_{ - \infty }^\infty  {\mathrm{WE}{z}(x,y) \cdot \hat \delta  \, dx \, dy} } \\
{\rm{with}}\quad \hat \delta  &= \delta [p - x \cos (\vartheta ) - y \cdot \sin (\vartheta )]\nonumber
\end{eqnarray}
\end{subequations}
%
Here $p$ denotes the sampling parameter for parallel scanning in  the $(x,y)$ plane in the direction normal to  $(\cos(\vartheta), \sin(\vartheta)) $. \\

Finally, the $RT\{SPP\} $ and $WE\{SPP\} $ were scaled by the refractive index of the SPP and WE  material (Al), resulting in a 2-d phase front behind the SPP and WE. 
These phase modulated partial waves from path I and path II interfere in front of or in the  crystal  (A)  form with  excited wave fields of (A)  a Moiré-pattern \cite{Bonse1969,Hart1975}.

For the calculation of the interference pattern we used the same formula as used in neutron interferometry \cite{RauchWerner2000} and thus Eq.(2) in  \cite{Sarenac2016}, however \emph{without} the $ q\phi $ term in the cosine function: 
\begin{equation}
I(u,v) \propto {\rm{ }} \cos{\rm{\big[}}{\phi _0} + 2\pi ({D_\mathrm{SPP}}(u,v) + {D_\mathrm{WE}}(u,v)) \big]\
\label{Eq.:Iuv}
\end{equation}
Here $I(u,v)$ denotes the two-dimensional intensity distribution that would be measured  by a position sensitive neutron detector  with detector plane being normal to the O-beam and  in  the $(x,z)$-plane (see Fig.~\ref{fig:Rays in interferometer}), $ \phi_0$ is the phase of the partial wave, which can be tuned with a phase flag (as was used  in \cite{Clark},  Fig.~1), $D_\mathrm{SSP}(u,v)$  and 
$D_\mathrm{WE}(u,v)$  are the position  dependent thicknesses  at the point, where the neutron wave transmits the SPP and  WE, scaled by  $D_{\lambda} $.  We therefore calculate the point-to-point interference of partial waves I and II using  Eq.\ref{Eq.:Iuv}, assuming both  path length I and II of equal lengths. 
One can assume some difference in optical path lengths and a certain lateral coherence length $\sigma_x$ and longitudinal coherence length $ \sigma_y $, which allows for superposition of the partial waves. 

Our formula Eq.~\eqref{Eq.:Iuv} differs from Eq.~\eqref{Eq.:Sarenac}), i.e. Eq.(2) in \cite{Sarenac2016}, where a  beam with orbital angular momentum (OAM) of $ l\hbar $, with  $l = q$, was  assumed,  corresponding to a topological charge of the object beam:
\begin{equation}
\begin{aligned}
I &= \int_0^d \int_0^d |\Psi_r + \Psi_o|^2 \, dx' \, dy' \\
&= A + B \cos(k y - q \phi + \theta)
\end{aligned}
\label{Eq.:Sarenac}
\end{equation} 
In \cite{Sarenac2016}  $d$ was the beam size, and $d \gg \sigma$, and $A$ and $B$  experimental constants, $ \theta \sim \phi_0 $ in Eq.(\ref{Eq.:Iuv}).

Taking into account the negligible vertical coherence length \( \sigma_z \), the resulting two-dimensional interference pattern can be constructed by stacking the interference images corresponding to each \((x, y)\) position. 
In order to compare our simulations with the experimental interference pattern of \cite{Sarenac2016},  we used the resolution of the neutron detector as discretization parameter in the $(x,z)$ - plane ($ 100\,\mu \mathrm{m} \times 100\,\mu\mathrm{m}$)
All calculations have been performed with MATLAB~\cite{Matlab}.   
\begin{figure}[hbtp]
\centering
\includegraphics[width=\linewidth]{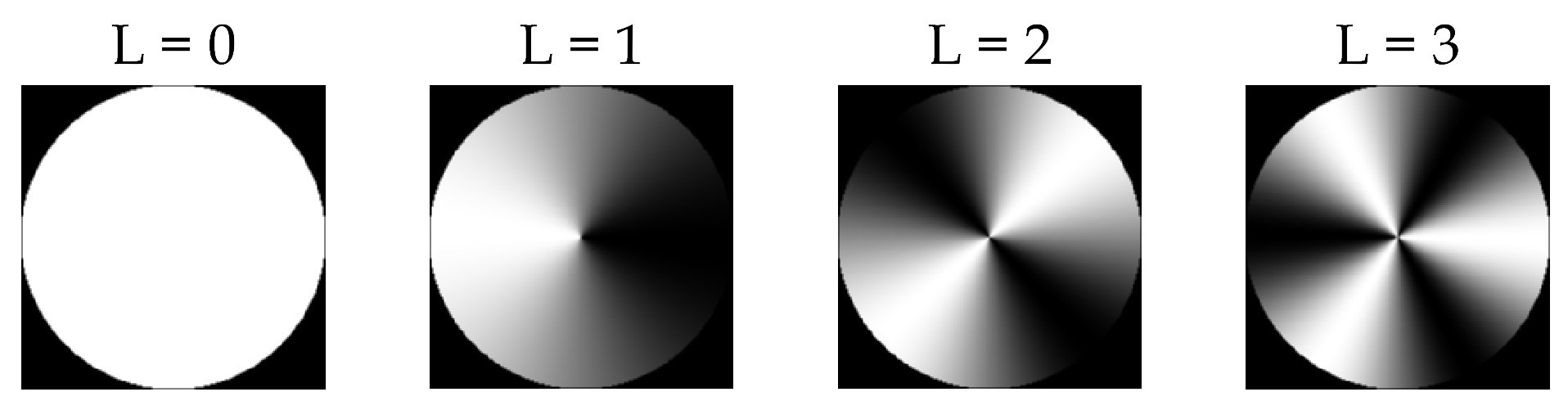} 
\includegraphics[width=\linewidth]{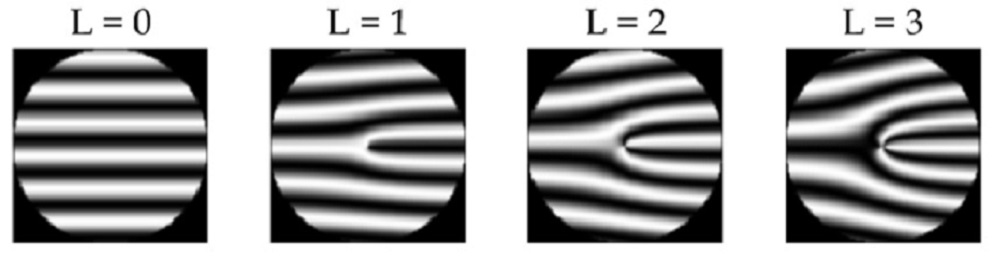}
\caption{Upper row: Interference pattern of spiral phase plates, only; bottom row: interference pattern from spiral phase plate  and phase wedge,  L  is the number of SPP in the interferometer,  all images were  calculated using Eq.(\ref{Eq.:Iuv}), also given in  \cite{RauchWerner2000}.}
\label{fig:interference_pattern}
\end{figure}

\begin{figure}[hbtp]
\centering
\includegraphics[width=\linewidth]{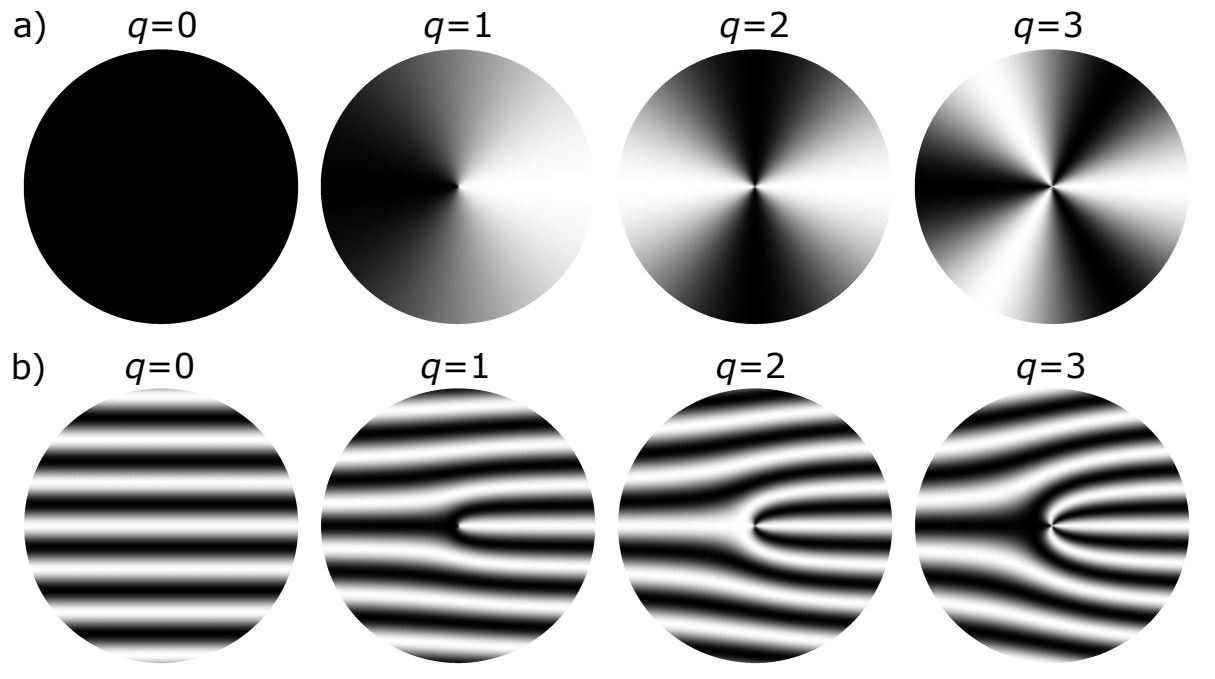}
\caption{Simulation  given in \cite{Sarenac2016}. Comparing  with our simulations  of Fig.~\ref{fig:interference_pattern}  the only difference is in the case where L= 0, which differs from the image for q = 0, see text.}
\label{fig:interference_pattern_Sarenac}
\end{figure}
The results are shown in Fig.~\ref{fig:interference_pattern} and are identical to the images in Fig.~\ref{fig:interference_pattern_Sarenac}  (Fig.~2 of \cite{Sarenac2016}). Any rotations between the simulations are due to a constant phase shift to the entire beam that may not have been included. For a similar reason, the only difference in our simulations is the case where $L = 0$ ($q = 0$), when both partial beams are empty.
This discrepancy arises because we used the camera position from Fig.~1(b) in \cite{Sarenac2016} for our calculations. The simulation for $q = 0$ in \cite{Sarenac2016} is valid only if the imaging camera is positioned at the location of the integration counter. \\

In the following, we present calculations using Eq.(3) from \cite{Sarenac2016}, which can also be found in \cite{Schnars1994}. For h(x,y), we use our interference patterns for L = 0 and 2 from Fig.~\ref{fig:interference_pattern}.   Eq.(3) from \cite{Sarenac2016} is given as 
\begin{equation}
\begin{aligned}
\Gamma (\nu ,\mu ) &= \frac{i}{{\lambda  d}}{e^{ - i \pi  \lambda  d ({\nu ^2} + {\mu ^2})}} \\
&\times F\left\{ {h(x,y) \cdot \exp ( - \frac{i\pi }{{\lambda  d}}({x^2} + {y^2})} \right\}
\end{aligned}
\label{Eq.:Eq._3_Sarenac}
\end{equation}
To calculate $ \Gamma(\nu,\mu)$, we used the discrete Fresnel transform described in \cite{Schnars1994}:
\begin{align}
\nonumber
\Gamma (m,n) & = \dfrac{i}{\lambda d}\exp \left[ { - i \pi   \lambda   d  \left( {\frac{{{m^2}}}{{{N^2} \cdot \Delta {x^2}}} + \frac{{{n^2}}}{{{N^2}  \Delta {y^2}}}} \right)} \right]  \\
\nonumber
 &\times \sum\limits_{k = 0}^{N - 1} {\sum\limits_{l = 0}^{N - 1} {h(k,l) \exp \left[ { - \frac{i\pi }{{\lambda   d}}\left( {{k^2}  \Delta {x^2} + {l^2}  \Delta {y^2}} \right)} \right]} }   \\
&\times \exp \left[ {i  2  \pi   \left( {\frac{{k  m}}{N} + \frac{{l  n}}{N}} \right)} \right]
\label{Eq.:discrete form FT}
\end{align}
with  $m = 0,1, \dots, N-1,\,  n = 0,1, \dots, N-1$,  $ \Delta x = 1$, $ \Delta y = 1$, $N = 100$ and  $ \lambda\, d = 70\, \mathrm{mm}^2$. 

\begin{figure}[hbtp]
\centering
\includegraphics[width=8cm]{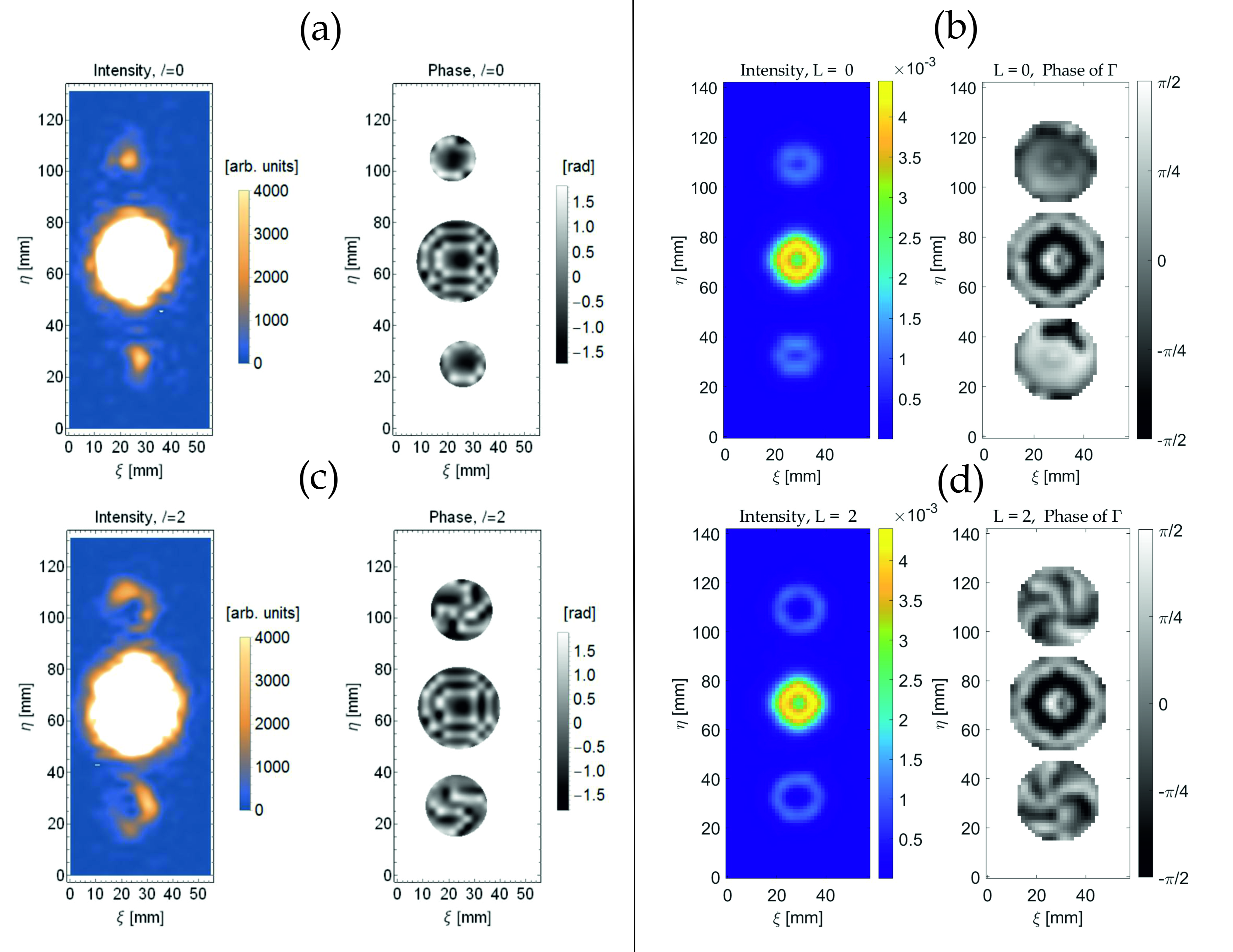} 
\caption{First Column:  Reconstructed intensity and phase images for $l=0$ (a) and $l=2$ (c) given in \cite{Sarenac2016}(Fig.~4), based on experimental neutron interferograms. Second Column:  Corresponding reconstructed  intensity and phase images for $L=0$ (b) and $L=2$ (d), based on calculated interferograms as shown in Fig.~\ref{fig:interference_pattern}. In both cases Eq.\eqref{Eq.:discrete form FT} was used for the reconstruction.} 
\label{fig:Phase reconstruction}
\end{figure}

With the discrete form of Eq.(\ref{Eq.:Eq._3_Sarenac}) from \cite{Schnars1994}, Eq.(\ref{Eq.:discrete form FT}), the intensity  $ |\Gamma|^{2} $ and the phase $ \tan^{-1}(\Im (\Gamma) / \Re (\Gamma) )$  of the interferograms $L = 0$ and $L = 2$ of Fig.~\ref{fig:interference_pattern} were calculated.
With $ \xi =  \nu\cdot \lambda  d$  and  $ \eta =  \mu\cdot \lambda  d$   (see Eq.(~\ref{Eq.:Eq._3_Sarenac})) and  $ \lambda \cdot $d = 70mm$^{2}$  it is not surprising to get (nearly) the same results (see Fig.\ref{fig:Phase reconstruction}).\\

{Our results (b) and (d) and the corresponding images from \cite{Sarenac2016}, (a) and (c),  are shown in Fig.\ref{fig:Phase reconstruction} for comparison. The reconstructed intensities for  l = 0 and l = 2  are different to L = 0 and L = 2,   however, the phases of the upper and lower satellites are similar in (a) and (b), as well as in (c) and (d) (Fig.~\ref{fig:Phase reconstruction}).
This behavior is expected, as the interference pattern is not a result of Fresnel scattering, which is a prerequisite for holographic recordings. Fresnel scattering describes wave propagation in the near field, which was not the case in the  experiments \cite{Sarenac2016}.
}

\section{Conclusions}

In this article, we have shown that the interference patterns described in \cite{Sarenac2016} as neutron holograms are a misinterpretation of normal interference patterns for several reasons:

(a) Each interference pattern recorded with a Mach-Zehnder (MZ) type crystal interferometer is measured as a simple scalar phase image of ordinary neutrons. This occurs because, first, the lateral coherence lengths are far too small and, second, they differ by two orders of magnitude. The observed pattern represents point-to-point interference between the two partial beams, I and II.\

(b) The realization of an MZ interferometer for neutrons is always associated with a Moiré pattern by virtue of dynamical diffraction in the Laue geometry. The phase-shifted waves of partial beams I and II generate an interference pattern that is superimposed on the reflection plane (and only on the reflection plane) of the analyzer crystal, producing a Moiré effect.\

(c) The conditions for neutron holography presented  in \cite{Sarenac2016}  are not met.   A certain coherence is important for holography, which makes it possible for an object beam and a reference beam to interfere and generate an interference pattern that contains both intensity and phase.
The holography of a spiral phase plate, as used  in \cite{Sarenac2016},  requires Fresnel diffraction and coherent illumination to achieve superposition with the reference beam. The condition to be met is, that the so-called Fresnel number $ F > 1 $. We have shown that, even with the optimal configuration of the wedge and object in the interferometer,  $F <  1$, which does not fulfill the condition for Fresnel scattering.\

(d)  In terms of spatial resolution,  the detector must have sufficient spatial resolution to accurately detect the fine interference fringes caused by the interaction of object and reference waves, which is definitely not the case in \cite{Sarenac2016}. \\

Finally, we want to carry out a short calculation to check whether a neutron OAM can be detected at all with a neutron interferometer as published in \cite{Sarenac2016}. 
Since the azimuthal phase is imposed only on neutron wave packets within a maximum of one coherence length from the spiral axis, we make the unrealistic assumption that $ \sigma_x  = \sigma_z = 1 \,\mathrm{\mu m}  $  (noting that  in reality  $ \sigma_z \sim  0.05 \, \mathrm{\mu m}$),  and consider a 50 fold  higher  neutron flux in the interferometer (as was in \cite{Sarenac2016}),  $10^{3} \, \mathrm{n / cm^{2}s} $. In this case, only 4 - 5 neutron wave packets would get a spiral wave front in 100 hours of experiment time and generate a neutron angular momentum.  With the neutron counting rate of  20 neutrons/$ \mathrm{cm}^{2}\mathrm{s}$  as reported in \cite{Sarenac2016},  it is impossible to detect the neutron orbital angular momentum. The patterns described in \cite{Sarenac2016} can only be explained as classical point-to-point neutron interferences of neutron waves via path (I) and path (II). The question is whether one can derive an orbital angular momentum (OAM) with characteristic topological charges from the ordinary interference patterns measured there.
\\

To summarize, we are strongly convinced that the neutron interference patterns published in \cite{Sarenac2016} cannot be considered neutron holograms. There are simply too many physical arguments against this interpretation: the crystal neutron interferometer is incorrectly treated as a Mach-Zehnder light interferometer, the lateral coherence lengths differ by approximately two orders of magnitude and are too small, and, moreover, the condition of Fresnel scattering on the object - essential for holography - is not satisfied.\\

\textbf{Author Contributions} W.T. had the idea,  F.H. calculated the spiral phase plate and the phase wedge, F.H. and W.T. calculated the interference patterns, M.S. and W. T. contributed the mathematics concerning dynamical theory and neutron interferometry, W. T., F.H., T.J. and I.B. contributed to neutron orbital angular momentum and holography, W.T.,  F.H.,  M.S. and T.J. wrote the manuscript. \\

\textbf{Author Information}  The authors declare that they have no competing financial interests.\\

\textbf{Code availability statement} We used MATLAB~\cite{Matlab}   to calculate the interferograms  presented in Figures 10, 11 and 12. 
{{The use of a software package does not imply recommendation or endorsement by the National Institute of Standards and Technology.}\\
MATLAB-scripts can be accessed via the public  Gitlab-repository} ( \url{..  will_be_submitted_later ...  }) \\

\section*{Acknowledgments}
One of us (W.T.) thanks the Helmholtz Zentrum Berlin  for its hospitality. \\


\bibliographystyle{apsrev}
\bibliography{holography}

\end{document}